# Metallic solid-state hydrogen storage crystals achieved through chemical precompression under ambient conditions

Baiqiang Liu[1], Chenxi Wan[1,2], Rui Liu[1], Zhen Gong[1], Jia Fan[1,2], Zhigang Wang[1,2,3,*]

[1] Key Laboratory of Material Simulation Methods & Software of Ministry of Education, College of Physics, Jilin University, Changchun 130012, China.
[2] Institute of Atomic and Molecular Physics, Jilin University, Changchun 130012, China.
[3] Institute of Theoretical Chemistry, College of Chemistry, Jilin University, Changchun 130023, China.
[*] e-mail: wangzg@jlu.edu.cn (Z. W.)

**ABSTRACT**

Improving hydrogen storage density is essential for reducing the extreme conditions required in applications such as nuclear fusion. However, the recognition of metallic hydrogen as the "Holy Grail" of high-pressure science highlights the difficulty of high-density hydrogen aggregation. Here, we report a solid-state crystal $H_9@C_{20}$ formed by embedding hydrogen atoms into $C_{20}$ fullerene cages and utilizing chemical precompression, which remains stable under ambient pressure and temperature conditions and exhibits metallic properties. This precompression effect is reflected in the formation of C–H bonds within the cage and C–C bonds between cages, resulting in the transformation of all C atoms from $sp^2$ to $sp^3$ hybridization with inward and outward distortions, while promoting delocalized multicenter bonding within the $H_9$ aggregate. In particular, the hydrogen density inside the $C_{20}$ cage exceeds that of solid hydrogen, achieving a uniform discrete distribution with $H_9$ as monomers. Further study reveals that filling hydrogen molecules into voids between $H_9@C_{20}$ primitive cells can increase hydrogen content while maintaining structural stability, forming a solid–gas mixed hydrogen storage crystal. Our findings provide a basis for developing high-density hydrogen storage materials under ambient conditions.

**INTRODUCTION**

The efficient storage of hydrogen has long been an important and even central focus in fields such as physics, chemistry, materials, and energy science.[1-4] In particular, since metallic hydrogen has been theoretically predicted,[5] its extremely high atomic density is considered to hold great potential for energy storage and as a nuclear fusion fuel.[6,7] However, achieving the pressures of hundreds of GPa required for its synthesis has long been extremely challenging,[8] and no universally recognized sample has yet been obtained.[9] More crucially, hydrogen energy utilization, exemplified by nuclear fusion, typically operates under ambient pressure conditions.[10,11] This clearly indicates that achieving the highest possible hydrogen storage density at low or even ambient pressures represents an important direction for future development.

Actually, to overcome the bottleneck of high-pressure conditions, researchers have developed a synthetic pathway that utilizes the intrinsic pressure of chemical bonds to achieve chemical precompression,[12] enabling the metallization of hydrogen at lower pressures.[13,14] Representative systems such as $H_3S$ and $LaH_{10}$ demonstrate metallization at approximately 150-200 GPa,[15,16] far below those required for pure hydrogen, yet still too high for practical applications. Meanwhile, since these structures generally contain elements with higher Z than hydrogen, which often leads to energy losses in hydrogen utilization,[17,18] their potential for practical applications is further reduced. This is also one of the important reasons why low-Z materials, such as hydrocarbon foams or high-density carbon shells, are commonly selected as fuel targets in inertial confinement fusion devices, including the National Ignition Facility.[19,20] Furthermore, for the application of hydrogen energy, such as in nuclear fusion, improving the absorption efficiency of externally injected energy is equally important,[21,22] and this usually requires high-density hydrogen aggregates to exhibit a uniform discrete distribution behavior. Notably, ordered hydrogen distribution via organic frameworks and fullerene surfaces offers a bottom-up perspective for this goal.[23-27] Based on these existing studies, developing hydrogen storage crystals with low-Z elements to achieve high storage density at low pressure and ensure a uniform discrete distribution of hydrogen aggregates is of great significance.

As potential containers capable of hosting hydrogen, nanoscale confinement structures may offer unique advantages for achieving efficient hydrogen aggregation. It has been shown that $C_{80}$ can stabilize molecular configuration that is unstable under ambient pressure and enhance intramolecular bond strength.[28] Similarly, hydrogen molecules have been embedded within $C_{60}$.[29] However, since the cavity of $C_{60}$ is much larger than the size of hydrogen molecules, the interaction with the cage wall is weak, making it difficult for hydrogen molecules to dissociate or form high-density aggregates, which is a key prerequisite for metallic hydrogen.[30] Meanwhile, embedding multiple hydrogen molecules can increase the internal pressure, ultimately destabilizing and rupturing the fullerene cage.[31,32] These studies seem to imply that by utilizing strong bonding mechanisms,[33,34] atomic hydrogen can be embedded into smaller fullerene molecules, potentially achieving effective hydrogen aggregation under mild conditions. As the smallest fullerene, $C_{20}$ has always been challenging to synthesize, and its relatively low thermodynamic stability poses obstacles for practical applications.[35,36] However, it is worth noting that recent studies have shown that embedding atoms into $C_{20}$ for strong bonding can indeed improve its stability.[37] Moreover, as fullerene crystals have been widely predicted and even experimentally synthesized,[38-40] the feasibility of using $C_{20}$ as a hydrogen storage unit to construct fullerene-based hydrogen storage crystals is greatly increased. Although no related studies have yet been found so far, this is what the present study aims to address.

Here, we report a solid-state three-dimensional hydrogen storage crystal $H_9@C_{20}$, which

achieves efficient hydrogen storage through chemical precompression and remains stable under ambient pressure and temperature conditions. Specifically, the formation of intra-cage C–H and inter-cage C–C bonds drives the transformation of all C atoms from $sp^2$ to $sp^3$ hybridization, inducing a cooperative distortion of the $C_{20}$ framework that is crucial for the crystal's stability. More importantly, this precompression effect suppresses hydrogen diffusion, leading to a hydrogen density exceeding that of solid hydrogen and enabling delocalized multicenter bonding, achieving a uniform discrete distribution with $H_9$ as the unit. Further electronic structure analyses reveal that $H_9@C_{20}$ crystal is metallic, defying the view that hydrogen is normally insulating or semiconducting under ambient conditions. Moreover, the voids between $H_9@C_{20}$ primitive cells can be further filled with hydrogen molecules to enhance the storage density. Overall, our findings lay the foundation for achieving high-density hydrogen storage crystals under ambient conditions.

**RESULTS AND DISCUSSION**

To achieve the crystallization of high-density hydrogen aggregate within $C_{20}$ fullerenes, we perform extensive structural modeling and periodic optimization. Eventually, a stable $H_9@C_{20}$ crystal is obtained, in which the $H_9$ aggregate is embedded in the $C_{20}$ cage. As shown in Figure 1a, within the primitive cell, eight hydrogen atoms bond to the $C_{20}$ cage via C–H bonds, while the remaining hydrogen atom resides at its center. The crystal belongs to a cubic structure in the highly symmetric space group *Pm-3m* (No. 221), with an equilibrium lattice constant of 5.52 Å (detailed structural parameters are provided in Table S1). It is worth noting that the isolated $C_{20}$ fullerene consists of C atoms with $sp^2$ hybridization and exhibits an approximately spherical structure,[35,41] making it difficult to stack further into a stable three-dimensional crystal. Importantly, the introduction of hydrogen atoms alters this situation by inducing the formation of C–H bonds, causing some C atoms to contract inward, while the remaining ones expand outward due to the inter-cage C–C bonding. As a result, the C atoms, originally dominated by $sp^2$ hybridization, undergo an overall transformation to $sp^3$ hybridization, endowing the $C_{20}$ cage with the ability to form three-dimensional covalent connections. Meanwhile, the hydrogen atoms are stabilized within the lattice through chemical bonding, thereby forming a solid-state hydrogen storage crystal. The change in hybridization type of C atoms on $C_{20}$ is also supported by orbital composition analysis (Table S2). Where the 2s orbital of each C atom undergoes compositional mixing with $2p_x$, $2p_y$, and $2p_z$ orbitals, leading to the formation of four bonds with $sp^3$ hybridization characteristics. Thus, adjacent $H_9@C_{20}$ primitive cells in the supercell are interconnected via C–C bonds (Figure 1b). Such covalent connections endow the crystal with remarkable structural stability.

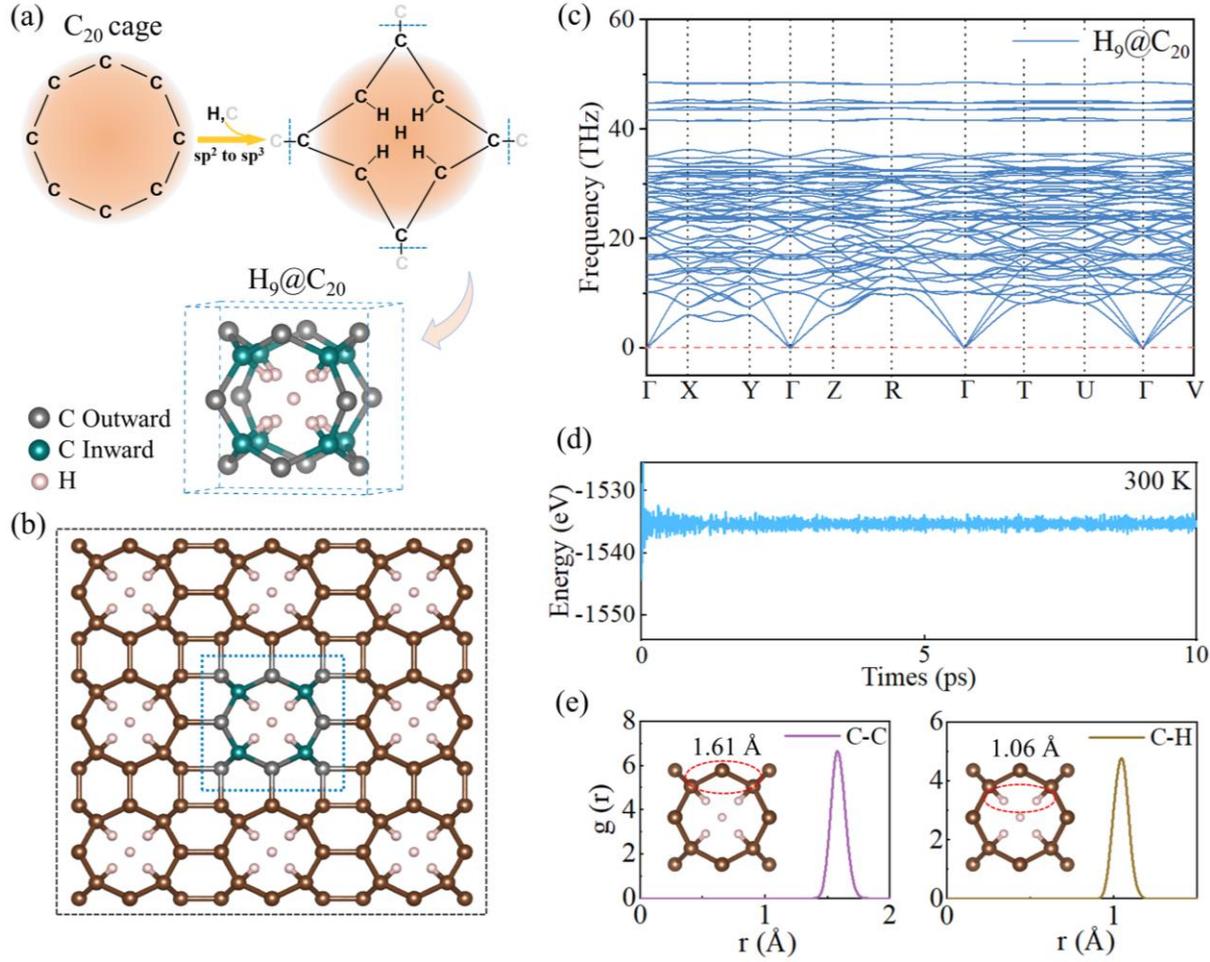

**Figure 1.** Optimized geometric structure and stability verification of the $H_9@C_{20}$ crystal. **(a)** Schematic diagram of hydrogen atoms embedded in $C_{20}$ fullerene for crystallization. Here, the formation of C–H bonds within the cage and C-C bonds between cages together causes partial inward contraction and outward expansion of C atoms on $C_{20}$. The blue-green C atoms represent inward contraction and the grey ones correspond to outward expansion. **(b)** Supercell of the $H_9@C_{20}$ crystal, with the primitive cell indicated by the blue dotted lines. Adjacent primitive cells are connected via covalent C–C bonds. **(c)** Phonon dispersion curves indicating the dynamical stability of the crystal. **(d)** Total energy variation over time at 300 K from AIMD simulations. **(e)** RDF of C-C atomic pairs and C-H atomic pairs from the AIMD simulations trajectories ranging from 0 to 10 ps. The dotted circle indicates the bond lengths of C-C and C-H bonds.

Remarkably, the optimized structure remains stable under ambient pressure, which sharply contrasts with the conventional view that hydrogen can only achieve effective aggregation under high pressure. This can be attributed to the complex interactions within the crystal composed of hydrogen and fullerene cages, thereby eliminating the need for external pressure. Interestingly, whereas earlier studies concluded that small fullerenes such as $C_{20}$ have difficulty in incorporating hydrogen due to size limitations and strong repulsion,[42] our results reveal differences between periodic crystal effects and the behavior of isolated molecules, highlighting the complexity of the hydrogen storage mechanism in crystals. These phenomena require in-depth analysis to be fully understood.

Based on the $H_9@C_{20}$ crystal structure, we evaluate its stability from dynamic, thermodynamic, and mechanical perspectives by calculating the phonon spectrum, conducting *ab initio* molecular dynamics (AIMD) simulations, and determining its elastic constants. Figure 1c shows the phonon dispersion curve in the first Brillouin zone. The absence of imaginary frequencies indicates that the structure is dynamically stable. From a thermodynamic perspective, we perform 10 ps AIMD simulations at 300 K, and no obvious fluctuations are observed in the total energy over time (Figure 1d). Furthermore, the radial distribution functions (RDF) show that both C–C and C–H bonds oscillate about their equilibrium positions (Figure 1e), which further confirms the thermodynamic stability at room temperature. To assess its mechanical properties, we calculate its elastic constants. For systems with cubic symmetry, mechanical stability requires satisfying the Born-Huang criteria ($c_{11} > 0$, $c_{44} > 0$ and $c_{11} + 2c_{12} > 0$).[43] For $H_9@C_{20}$, the calculated elastic constants are $c_{11}$ = 673.1 GPa, $c_{12}$ = 35.6 GPa, and $c_{44}$ = 152.1 GPa, all of which meet the above criteria, thus confirming its mechanical robustness. Collectively, these indicators jointly demonstrate that the three-dimensional $H_9@C_{20}$ crystal has excellent stability under ambient conditions. Furthermore, given the three-dimensional covalent connectivity of the $C_{20}$ cage structure, the material is expected to exhibit excellent hardness. The calculation results show that its bulk modulus is 248.09 GPa and the shear modulus is 205.51 GPa (Table S3), indicating strong resistance to uniform compression and shear deformation. In the Vickers hardness test, $H_9@C_{20}$ achieved a hardness of 33.02 GPa. Although this value is lower than that of diamond (~92 GPa),[44] it exceeds that of conventional hard covalent materials such as SiC (28 GPa).[45] These results indicate that the $H_9@C_{20}$ crystal not only exhibits remarkable stability but also holds the potential to be suitable for applications requiring high hardness and deformation resistance.

After confirming the stability of the crystal, we analyze its bonding mechanism using the electron localization function (ELF) and crystal orbital Hamilton population (COHP) methods. As shown in Figure 2a, both the C–C bonds between primitive cells and the C–H bonds within the primitive cell exhibit electron localization, thereby confirming the formation of covalent bonds. This bonding feature agrees with the chemical precompression theory,[12] wherein the internal compressive effect generated by chemical bonding enables hydrogen atoms to form a dense arrangement similar to that under high-pressure conditions, and enhance the overall structural stability. Notably, appreciable multicenter electron delocalization is clearly observed between the central hydrogen atom and the eight hydrogen atoms that are bonded to the C atoms ($H_8$). Continuous yellow-green regions (ELF ~ 0.5) appear between the hydrogen atoms, in contrast to the typical deep-blue, isolated interatomic regions. Such a distribution indicates that the electron cloud extends across atomic boundaries, forming a delocalized electronic state. This feature resembles the electronic behavior of metallic hydrogen,[5,46] further suggesting that the chemical precompression effect enables hydrogen atoms to effectively aggregate. Additionally, COHP analysis of the C–C bonds in the $H_9@C_{20}$ crystal shows that, below the Fermi level, the bonding orbitals are stronger than the antibonding ones, with an integrated COHP (ICOHP) value of -4.45 eV, indicating the formation of covalent bonds (Figure 2b). Similarly, the C–H bonds also exhibit strong bonding characteristics, with an ICOHP value of -3.82 eV. Between hydrogen atoms, bonding orbitals dominate, indicating an attractive interaction (ICOHP = -0.51 eV) that facilitates electron delocalization across atomic boundaries, thereby forming a metal-like hydrogen state.

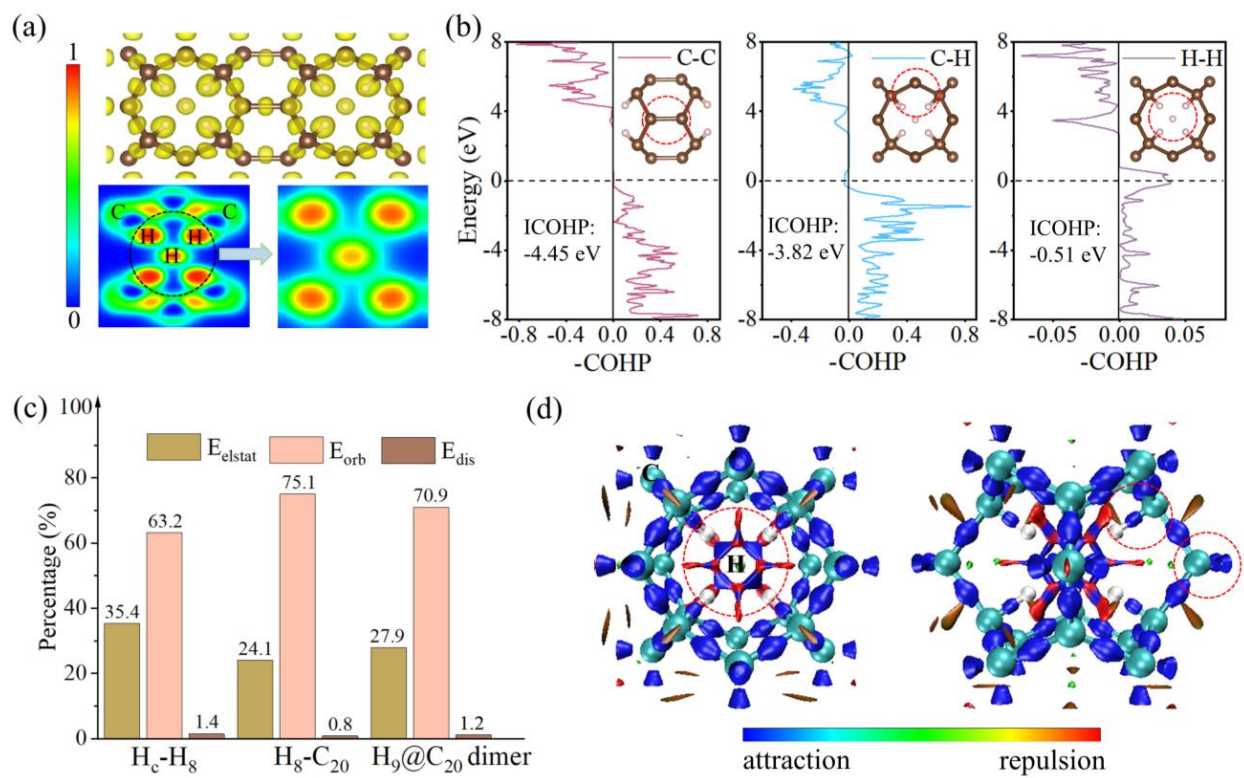

**Figure 2.** Interaction analyses within and between $H_9@C_{20}$ primitive cells. **(a)** ELF map. ELF values close to 0 (blue) represent regions of delocalized electrons, values around 0.5 (green) correspond to uniform electron gas, and values close to 1 (red) indicate highly electron localization. **(b)** COHP and ICOHP analyses. C–C and C–H interactions represent bonding between and within the primitive cells, respectively. H–H interactions refer to those between the central hydrogen atom and the hydrogen atoms bonded to the C atoms. **(c)** Energy decomposition analysis (EDA). Here, $H_c$ represents the central hydrogen atom, and $H_8$ represents the eight hydrogen atoms that are bonded to the C atom. $E_{elstat}$ represents electrostatic interaction, $E_{orb}$ represents orbital interaction, and $E_{disp}$ represents dispersion interaction. **(d)** IRI analysis. The isosurfaces in blue, green, and red represent attractive, van der Waals (vdW), and repulsive interactions, respectively.

To quantitatively elucidate the contributions of different attractive interactions to the stability of the $H_9@C_{20}$ crystal, we conduct an energy decomposition analysis (EDA). As shown in Figure 2c, the orbital interaction dominates the attractive interaction between the $H_8$ and the $C_{20}$ cage, accounting for 75.1%, which is higher than the contributions from electrostatic interaction (24.1%) and dispersion interaction (0.8%). This result confirms orbital hybridization and covalent bond formation between C and hydrogen atoms. Moreover, orbital interaction also dominates the interactions between adjacent primitive cells (70.9%), consistent with the covalent crystal characteristics revealed by ELF and COHP analyses. Notably, the stability of the central hydrogen atom arises primarily from its orbital interaction with the $H_8$, contributing 63.2% of the total attractive interaction, which indicates orbital overlap between the hydrogen atoms. These conclusions are further supported by the results of the interaction region indicator (IRI) analysis. As shown in Figure 2d, an attractive interaction exists between the $H_8$ and the C atoms, and this interaction also exists between the $H_8$ and the central hydrogen atom. In addition, under the influence of chemical precompression, the $C_{20}$ cage framework becomes stable and exerts steric hindrance effect on its interior, which also promoting the stability of the central hydrogen atom.

Nevertheless, the stability of the central hydrogen atom is fundamentally determined by the orbital interactions between the hydrogen atoms.

It is worth mentioning that the density of the $H_9$ aggregate inside the $C_{20}$ cage is 0.29 g/cm³, which is higher than that of solid hydrogen (~0.086 g/cm³).[47] More importantly, this study is based on a fullerene structure and achieves a three-dimensional solid hydrogen storage crystal through chemical precompression. In this system, hydrogen atoms exhibit a highly aggregated state with characteristics similar to metallic hydrogen, which is distinct from previous approaches that relied on hydrogen adsorption on fullerene molecules.[25-27] This discovery may provide new opportunities for the utilization of hydrogen energy, including nuclear fusion reactions. In addition, the uniform discrete distribution of hydrogen fuel and the use of low-Z elements can potentially enhance fuel utilization efficiency and the absorption of externally injected energy.[10,20] In our work, although the hydrogen storage gravimetric density in this system does not reach that of some existing hydrogen storage materials due to remaining voids between $H_9@C_{20}$ primitive cells, under ambient conditions this system achieves a uniform discrete distribution as well as high-density encapsulation of hydrogen atoms within $C_{20}$ cage composed solely of low-Z element. This configuration offers enormous potential for improving fuel utilization efficiency and enhancing the absorption of external energy during fusion and other processes. Furthermore, it is evident that filling these voids can further enhance hydrogen storage density and improve related performance, which requires more extensive and in-depth exploration in multiple aspects in the future. Therefore, the crystal we design can serve as a fundamental conceptual model and become the basis for further development of hydrogen storage.

In order to elucidate the electronic structure properties of the $H_9@C_{20}$ crystal, high-precision band structure and density of states (DOS) calculations are performed. As shown in Figure 3a, the valence band of the $H_9@C_{20}$ crystal crosses the Fermi level, indicating metallic properties under ambient conditions. Further DOS analysis supports this conclusion. The total DOS (TDOS) near the Fermi level is continuously distributed, satisfying the electron state density condition for metal conduction (Figure 3b). Projected DOS (PDOS) analysis reveals that the electronic states near the Fermi level are primarily derived from C atoms, while hydrogen atoms also exhibit metallic behavior. Notably, this finding contradicts the conventional view that hydrogen typically behaves as an insulator or semiconductor in the solid state,[48,49] highlighting that chemical precompression effect can endow hydrogen with unique electronic behavior. Furthermore, since the central hydrogen atom in the $H_9@C_{20}$ structure does not directly form a chemical bond with the C atoms, we also perform a comparative study of the $H_8@C_{20}$ crystal. The phonon dispersion curve in the first Brillouin zone shows no imaginary frequencies (Figure S1), indicating that this structure is also dynamically stable. Further analysis of the band structure and DOS shows that $H_8@C_{20}$ exhibits semiconducting behavior (Figures S2 and S3). This comparison highlights the crucial role of the central hydrogen atom in enabling the crystal to exhibit metallic properties.

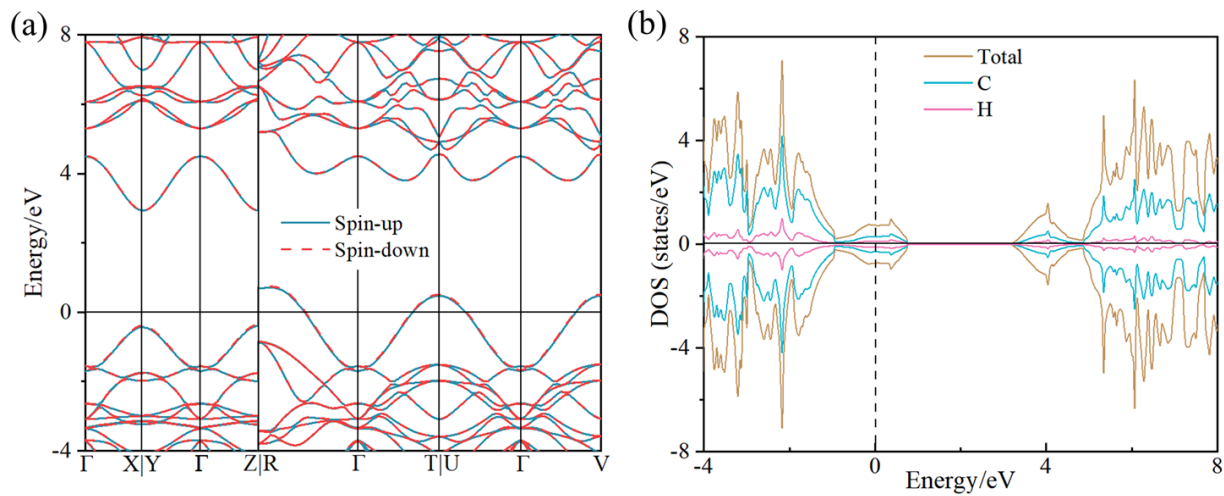

**Figure 3.** Electronic structure properties of the $H_9@C_{20}$ crystal. **(a)** Band structure of the $H_9@C_{20}$ crystal. The $E_F$ denoted by the horizontal dashed line is set to zero. The high symmetry k points are Γ(0, 0, 0), X(0.5, 0, 0), Y(0, 0.5, 0), Z(0, 0, 0.5), R(0.5, 0.5, 0.5), T(0, 0.5, 0.5), U(0.5, 0, 0.5), and V(0.5, 0.5, 0). **(b)** DOS analysis.

To elucidate the formation mechanism of $H_9@C_{20}$ under ambient conditions, we analyze the orbitals contributing to the interaction between $H_9$ and $C_{20}$. An $H_9@C_{20}$ fragment with $C_{4v}$ point group symmetry is extracted from the $H_9@C_{20}$ crystal, and the composition of its occupied orbitals is examined. As shown in Figure 4a, the energy level diagram reveals that the orbitals of the $H_9$ aggregate play a crucial role in bond formation within the $H_9@C_{20}$ fragment. Specifically, the A1 orbitals of the $H_9$ aggregate interact with the orbitals of the $C_{20}$ structural framework in the low-energy region, forming the A1 bonding orbitals of $H_9@C_{20}$. This orbital coupling enhances the bonding stability of the $H_9@C_{20}$ fragment. Moreover, the E orbitals of the $H_9$ aggregate also participate in bonding interactions with the $C_{20}$ framework orbitals in the higher-energy region, further promoting the formation of $H_9@C_{20}$ (the details in Table S4). Overall, these results demonstrate that the orbitals of the $H_9$ aggregate play a key role in the formation of the $H_9@C_{20}$ monomer.

Next, to more accurately examine the electronic behavior of the $H_9$ aggregate within the $C_{20}$ cage, we perform local molecular orbital (LMO) analysis to reveal its electronic distribution, thereby determining how it participates in chemical bonding. As shown in Figure 4b, LMOs based on the $H_9$ aggregate present delocalized multicenter bonding, consistent with the delocalized multicenter electron density revealed by ELF and previous findings on stability via such bonding.[50] Subsequently, we select the largest diagonal cross- section defined by the central hydrogen atom and four adjacent edge hydrogen atoms to analyze its electron density distribution. The critical electron density corresponding to the van der Waals radius of hydrogen was first determined, and integration of the electron density over all hydrogen atoms yielded the corresponding critical value $\rho(r_{vdW}) = 0.0158$ e·Å$^{-3}$. This critical value is used as a standard for quantifying the degree of electron delocalization in the $H_9$ system and is presented in Figure 4c as the electron distribution contour. It can be observed that the $H_9$ aggregate exhibits a continuous electron density distribution, with the electron cloud spanning the boundaries of multiple hydrogen atoms and forming clear delocalized multicenter feature, consistent with the LMO results above.

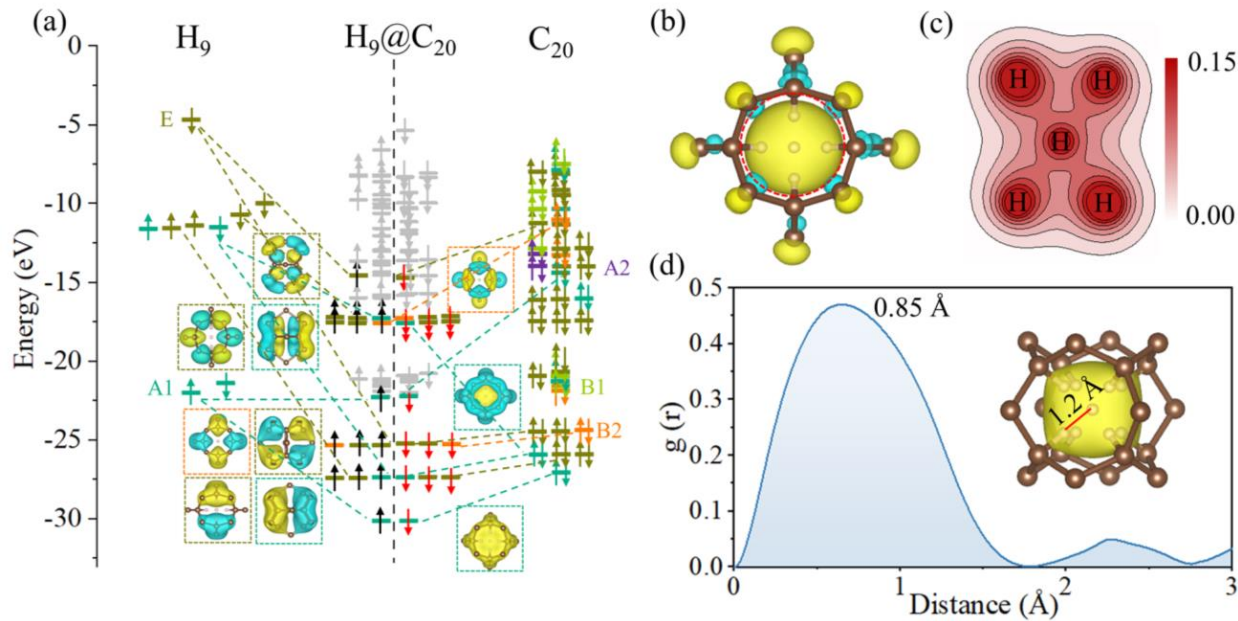

**Figure 4.** The energy level diagram and electronic properties. **(a)** The $H_9@C_{20}$ fragment exhibits $C_{4v}$ point group symmetry. Except for the gray-labeled orbitals, which originate from $C_{20}$, the remaining orbitals are mainly derived from both $H_9$ and $C_{20}$. **(b)** LMOs that localize on $H_9$ aggregate. **(c)** Electronic density analysis. The cross-section is taken in the plane defined by the central hydrogen atom and four diagonal hydrogen atoms. **(d)** RDF of the delocalized orbital in the $H_9$ system. In the RDF analysis, spherical integrals are evaluated radially from the system center.

Furthermore, to quantitatively characterize the spatial extent of this delocalized orbital, we calculate its RDF. Figure 4d shows that its main RDF peaks fall within the 0-1.8 Å range, with the approximately 1.2 Å distance between the central hydrogen and the top hydrogen lying within this interval, thereby further supporting the conclusion that delocalized multicenter bonding exists in the system. More importantly, the distance from the central hydrogen atom to the vertex hydrogen atoms is noticeably smaller than the H–H separations in some metal hydrides and in hydrogen under high pressure, [51,52] indicating that the chemical precompression effect can cause the $H_9$ aggregate to exhibit a highly compressed state under ambient conditions. In addition, unlike the previously reported cubic $C_{20}$ crystal,[53] our study reveals that the embedded hydrogen atoms can induce a cooperative deformation of the isolated $C_{20}$ fullerene, thereby forming a densely packed solid-state hydrogen storage crystal under ambient conditions. Overall, the combined analyses of LMO, spatial electron density, and RDF systematically and precisely depict the delocalized multicenter properties of the $H_9$ aggregate within the $C_{20}$ cage.

In terms of preparation methods, although the direct synthesis of such a high-density hydrogen storage crystal at ambient pressure is highly challenging, a feasible strategy may involve high-pressure synthesis followed by stabilization under ambient conditions. Previous experiments have successfully synthesized carbon nitride and boron nitride crystals using this method,[54,55] and even achieved the insertion of hydrogen atoms into the carbon nitride framework and the stable recovery under ambient pressure.[56] Of course, this also includes different practical possible preparation techniques. For example, in situ bottom-up growth can be achieved by precisely controlling the elemental stoichiometry under high can be achieved by precisely controlling the elemental stoichiometry under high pressure.[57,58] Alternatively, pre-synthesized $C_{20}$ cage structures can be combined with hydrogen, and crystal growth promoted through infiltration and related techniques under high pressure.[59] With the continued advancement of related experimental techniques, the synthesis of high-density hydrogen storage crystals proposed in this study is becoming increasingly promising.

Considering the presence of lattice voids in the $H_9@C_{20}$ crystal, we evaluate its potential for further enhancing hydrogen storage density. Specifically, without changing the framework of this solid hydrogen storage crystal, we investigate the feasibility of filling different numbers of hydrogen molecules into the voids between $H_9@C_{20}$ primitive cells. After high-precision structural relaxation, the results show that the crystal can remain stable under ambient pressure until four hydrogen molecules are filled (detailed structural parameters are provided in Table S5).

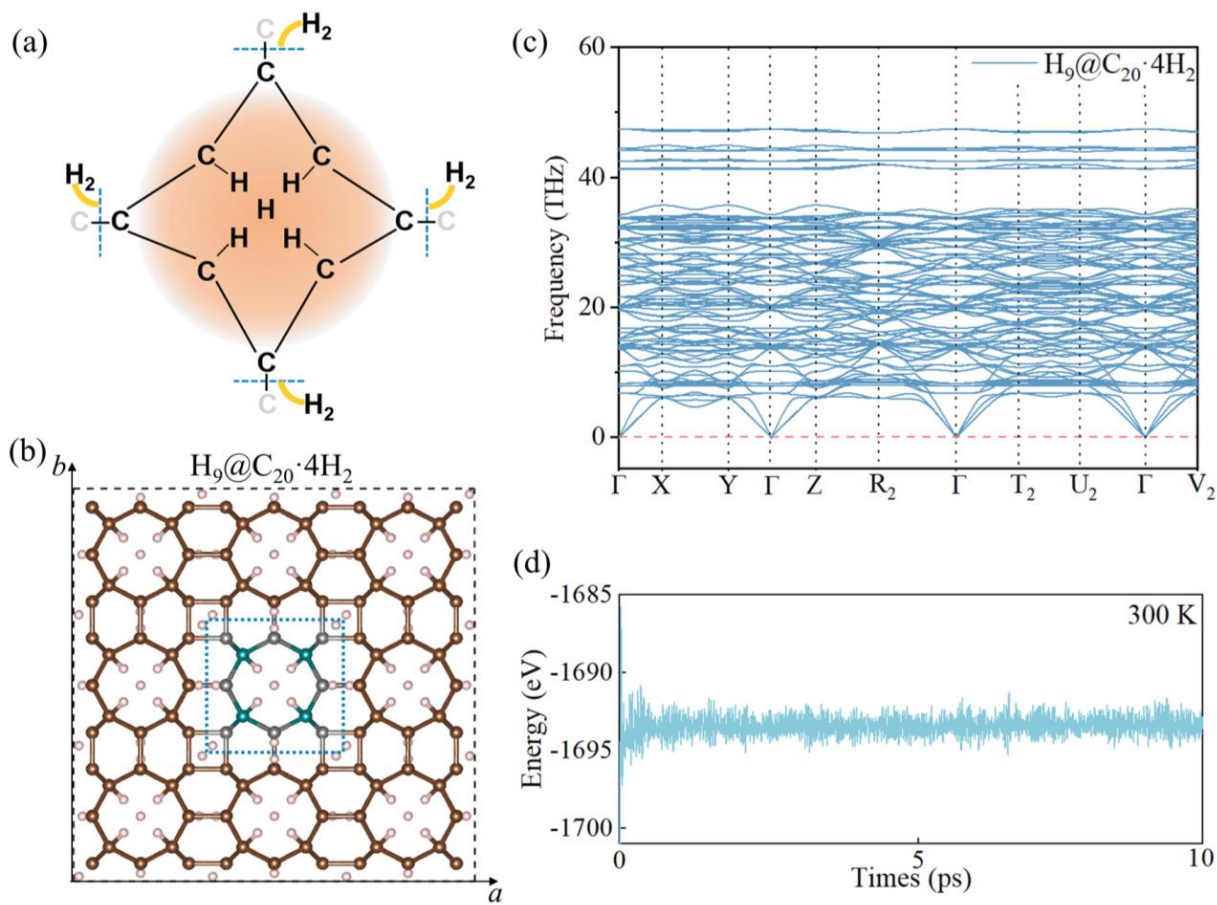

**Figure 5.** Optimized geometric structure and stability analysis of the $H_9@C_{20}\cdot 4H_2$ crystal. **(a)** Schematic diagram of hydrogen molecules embedded in the voids of $H_9@C_{20}$ crystal. Here, the blue dotted lines indicate the boundaries of the primitive cell. **(b)** Supercell of the $H_9@C_{20}\cdot 4H_2$ crystal. Four hydrogen molecules are stably located in the voids between the primitive cells, with one hydrogen molecule in each of the a, b, and diagonal directions, and another hydrogen molecule along the c direction (the details in Figure S4). **(c)** Phonon dispersion curves indicating the dynamical stability of the crystal. **(d)** Time evolution of the total energy at 300 K from AIMD simulations.

Taking the filling of four hydrogen molecules as an example, we analyze the dynamic and thermodynamic properties. As shown in Figure 5a and 5b, these four hydrogen molecules are stably distributed in the regions between adjacent primitive cells (detailed structural are provided in Figure S4). Further ELF analysis reveals that these embedded hydrogen molecules have not undergone any chemical bond rearrangement and exhibit no electronic overlap with $H_9@C_{20}$ (Figure S5). These results suggest that their stability primarily originates from van der Waals interactions. No imaginary frequencies are found in the phonon spectrum throughout the first Brillouin zone, confirming the dynamical stability of the $H_9@C_{20} \cdot 4H_2$ crystal (Figure 5c). Additionally, at 300 K, the 10 ps AIMD simulations reveal no obvious fluctuation in total energy over time, further confirming its thermodynamic stability (Figure 5d). In terms of hydrogen storage performance, this design increases the gravimetric density from 3.6 wt% in the base model to 6.6 wt%, while maintaining stability under ambient conditions. In summary, in the $H_9@C_{20} \cdot 4H_2$ crystal, the original solid-state hydrogen storage framework is completely preserved, and the bond in the filled hydrogen molecule is not broken either. Thus, a hybrid solid–gas hydrogen storage mechanism is achieved, which further provides a potential approach for designing high-density and stable hydrogen storage materials under ambient conditions.

**CONCLUSION**

Achieving efficient hydrogen aggregation under ambient conditions is crucial. We have proposed a strategy for embedding hydrogen atoms into $C_{20}$ fullerene cages to form a solid-state three-dimensional hydrogen storage crystal. Obviously, this hydrogen storage pathway offers four major advantages: stability under ambient pressure and temperature conditions, solid-state and high-density storage, uniform discrete distribution and low-Z composition. Through systematic analyses of the electronic structure and dynamics, the stability mechanism derived from chemical precompression has been elucidated. Specifically, the formation of C–H covalent bonds causes some C atoms to contract inward, while the formation of inter-cage C–C covalent bonds causes the remaining ones to expand outward. This collaborative effect plays a crucial role in the formation of the $H_9@C_{20}$ crystal. The resulting chemical precompression significantly shortens the H–H distances compared to those in conventional hydrides, leading to multicenter bonding within the $H_9$ aggregate, whose delocalized electronic behavior closely resembles that of metallic hydrogen under high pressure. Additionally, the primitive cells are interconnected through C–C covalent bonds, thereby endowing the crystal with excellent mechanical strength.

To elucidate the role of the central hydrogen atom, we have compared the electronic properties of $H_8@C_{20}$. This crystal exhibits semiconducting behavior, whereas $H_9@C_{20}$ displays metallic behavior under ambient conditions, demonstrating the critical role of the central hydrogen in triggering the semiconductor-to-metal transition. Moreover, based on the voids identified in the $H_9@C_{20}$ crystal, we also have constructed a mixed hydrogen storage system that combines atomic and molecular hydrogen. These findings contribute an important foundation for the design of high-density hydrogen storage materials that can remain stable under ambient conditions.

**ASSOCIATED CONTENT**

**Supporting Information.**

The Supporting Information is available free of charge via the Internet at xxx

Computational details, The structural parameters, orbital composition, and force constants of the $H_9@C_{20}$ crystal. It also provides the orbital energy levels of $H_9@C_{20}$ crystal, together with the phonon spectrum, band structure, and density of states (DOS) of the $H_8@C_{20}$ crystal, as well as the structural parameters, supercells, electronic structure of the $H_9@C_{20} \cdot 4H_2$ crystal.

**Author Contributions**

Z. W. initiated, designed and supervised the work. B. L. performed the theoretical simulations and calculations. The manuscript was written through contributions of all authors. All authors have given approval to the final version of the manuscript.

## ACKNOWLEDGMENT

This work was supported by the National Key Research and Development Program of China (No. 2024YFA1409900), the Science and Technology Development Program of Jilin Province of China (20250102014JC) and the National Natural Science Foundation of China (grant number 11974136).